# Patching Hele-Shaw cells to investigate the flow at low Reynolds number in fracture networks

**Pouria Aghajannezhad, Mathieu Sellier\*, and Sid Becker**
*Department of Mechanical Engineering, University of Canterbury, Christchurch, New Zealand*

## Abstract

This research has found a novel computational efficient method of modelling flow at low Reynolds number through fracture networks. The numerical analysis was performed by connecting Hele-Shaw cells to investigate the effect of the intersection to the pressure field and hydraulic resistance for given inlet and outlet pressure values.  In this analysis, the impact of intersecting length, intersecting angle and fracture aperture on the fluid flow was studied. For this purpose, two models with different topologies were established. The Hele-Shaw simulation results for hydraulic resistance, pressure and velocity agreed well with results obtained by solving the full Navier-Stokes equations (NSE). The results indicated an approximately linear relationship between intersection length and hydraulic resistance. Specifically, an increase in the intersection length increases the flow rate and as a result, the pressure along the intersection length decreases. The error associated with employing the Hele-Shaw approximation in comparison with NSE is less than 2%. All Investigations were performed in the Reynolds Number range of 1-10.

## Nomenclature

| | | |
|---|---|---|
| $\rho$ | Density $(kg/m^3)$ | |
| $\boldsymbol{u}$ | Velocity vector $(m/s)$ | |
| $P$ | Pressure $(Pa)$ | |
| $\mu$ | Viscosity $(Pa.s)$ | |
| $Q$ | Flow rate $(m^3/s)$ | |
| $Q_{H\_S}$ | Evaluated flow rate by Hele-Shaw approximation $(m^3/s)$ | |
| $Q_{N\_S}$ | Evaluated flow rate by Navier Stokes equations $(m^3/s)$ | |

\* Corresponding author.
*E-mail address:* mathieu.sellier@canterbury.ac.nz (M. Sellier).

| | |
|---|---|
| $h$ | Fracture aperture $(m)$ |
| $w$ | Fracture depth $(m)$ |
| $L$ | Fracture length $(m)$ |
| $\Delta P$ | Global pressure difference $(Pa)$ |
| $R$ | Hydraulic resistance $(Pa.s^{-1}.m^{-3})$ |
| $i$ | Intersection length $(m)$ |
| $\theta$ | Intersection angle |
| $U_{av}$ | Average flow velocity along a fracture |
| $A$ | Cross-section area $(A = w \times h)$ $(m^2)$ |

# 1. Introduction

From a geotechnical point of view, fractures play an essential role in groundwater fluid movement. Fractures allow the transport of geothermal energy, liquid fossil fuels and groundwater(National Research Council (U.S.). Committee on Fracture Characterization and Fluid Flow. 1996). A better understanding of the flow in fracture networks will enable a better evaluation of the geothermal resource. Even though several studies have been dedicated to investigating fluid flow in fractures, the fracture networks with a complex topology still present a challenge and are as yet far from being fully understood(Lei et al. 2017; Yu et al. 2017).

To deal with fracture intersections in numerical evaluations, most previous studies have focused on traditional Navier-Stokes equations (NSE)(Liu et al. 2018; Liu et al. 2016; Nazridoust et al. 2006; von Planta et al. 2020; Zimmerman et al. 2004). Despite the extended use of the NSE for simulation of steady-state incompressible flow in rock fractures, it is still challenging and demands high computational resources for large scale fracture networks – usually involving hundreds or thousands of intersections and fractures(Brush and Thomson 2003; Frampton et al. 2019). For example, the effect of hydraulic head, surface roughness and intersecting angle of two fractures were studied by solving NSE(Li et al. 2016), but techniques to solve the NSE for discrete fracture networks (DFN) with a large number of fractures are impractical.

Many studies in the last ten years have focused on the linearization of the NSE which leads to the cubic law (CL) relation(Gaucher et al. 2015). The CL can be obeyed in the Stokes regime when the flow velocity is small (Witherspoon et al. 1980), but the simulation of fluid flow in real fractures by the CL results in flow rate overestimation(Lee et al. 2014). In recent years, the CL has received increasing attention as a promising alternative method for the prediction of the fluid flow through rock fractures. The key motivation is not only to provide a computationally efficient numerical model, but also to enhance our understanding of the fluid flow behaviour through several fractures with different geometrical and topological properties(Konzuk and Kueper 2004). For instance, in (Guo et al. 2020), the experimental results of fluid flow through five tensile artificial cylindrical fractures in parallel connection were compared to the approximation of modified CL which was relatively in good agreement. However, in this model, the effect of intersection angle and intersection length were not considered. (Wang et al. 2018) developed a modified CL for studying fluid flow behaviour through a single fracture intersection. They found that the modified CL can approximate the fluid flow behaviour with an average error of 4.7%; thus, the use of modified CL can effectively eliminate the main drawback of flow rate overestimation. Additionally, the randomness of fracture geometries can be taken into account by a modification of the CL (Wang et al. 2015).

In recent years, several authors have attempted to develop computationally efficient methods such as graph theory for modelling DFN (Djidjev et al. 2017; Ghaffari et al. 2011; Hyman et al. 2017). Graph theory is a mathematical tool for the investigation of a network with connecting points (Srinivasan et al. 2018). (Hobé et al. 2018)investigated optimization algorithms based on graph theory for overall flow rate estimation of constant aperture fractures, and found a 16.62% average error in comparison with direct numerical simulation. The aperture is typically defined as a gap between two parallel plates. Full details of a three dimensional model for studying fluid flow through a DFN with low permeability were determined; specifically, multilevel graph partitioning method was utilised, which is a heuristic approach for partitioning large graphs based on graph theory (Ushijima-Mwesigwa et al. 2019). However, they have only focused on approximating the overall flow rate in a large scale DFN without considering fracture details. Although graph theory is an interesting approach and is computationally efficient to approximate the overall flow rate, it fails to take into account the geometrical details of each fracture and their collective topology. In other words, it is impractical to use this method for studying geometrical effects such as the intersection length of fractures on fluid flow.

Surface roughness can dramatically affect fluid flow behaviour and induce turbulence (Zou et al. 2015). To model fluid flow in fractures and study the effect of roughness, previous researchers have performed simulations of fluid flow in rough surface fractures (Fan and Zheng 2013; Morteza Javadi1, Mostafa Sharifzadeh1,2, Kourosh Shahriar1 2014; Plouraboué et al. 1998; Wu et al. 2018). As reported in a numerical study by (Zhang et al. 2019), the relative aperture of a fracture and its discharge per unit width is usually reduced by rougher surfaces. Some studies have indicated that large roughness in the direction perpendicular to flow will reduce the flow rate under some conditions. This can cause not only turbulent flow, but also influence the rock wall temperature, which is an important parameter for geothermal energy generation (Foroughi et al. 2018; Huang et al. 2019; Zhao et al. 2013). The flow turbulence in a fracture network is also influenced by other factors such as the hydraulic head, aperture variation and fracture intersection (Li et al. 2016). However, it has been demonstrated that roughness can cause nonlinear fluid flow response at low Reynolds numbers (Lee et al. 2014), meaning that for a large aperture, even for $Re < 1$, a nonlinear flow response exists. Research has tended to focus on solving the NSE to study fluid flow behaviour in rough surface rock fractures rather than the establishment of a computationally efficient method. The additional problem is that solving the NSE for a DFN with a large number of fractures is impractical (Koyama et al. 2008).

Despite various studies on fluid flow behaviour in rock fractures, much still remains to be done to increase computational efficiency and allow the treatment of a large number of connected fractures – which naturally have varying intersecting lengths, angles and apertures. The ability to study large discrete fracture network would shed light on the effects of geometrical and topological parameters such as intersecting length, angles, apertures, etc … This assessment can be helpful for choosing the right parameters for modelling the fluid flow in DFN of the geothermal field or subsurface solute transport. To comprehensively study the independent role of fracture geometry and topology in a DFN, modifying the HS approximation to provide a quick but sufficiently accurate model of the flow would be of real use to the field in advancing the current understanding of fracture network flows.

Using simplifying assumptions, the NSE can be replaced by the Hele-Shaw approximation (HS)(Zhao et al. 2013), a less demanding and better conditioned governing equation. Solving the HS equation for rectangular cells can provide solutions for complex flow problems in less time compared to the NSE. Henry Selby Hele-Shaw attained the two-dimensional hydrodynamic phenomenon by developing an experimental study on rectangular parallel plates with a narrow gap (HELE-SHAW 1898). Until now, many researchers have used this phenomenon for explaining fluid flow in porous media. Several authors have conducted experimental studies on single Hele-Shaw cells and evaluated the effect of surface roughness on fluid flow (Alturki et al. 2014; Planet et al. 2011). Many attempts have been made to simulate two-phase fluid in single Hele-shaw cells (Eslami et al. 2020; Singh et al. 2020). However, studies of fluid flow in multiple rock fractures are scarce and the majority of studies have tended to focus on flow in single fracture rather than multiple fractures. Despite the computational efficiency of the approach, no one to the best of our knowledge has studied multiple fractures using Hele-Shaw patching method.

The main objective of this paper is to develop computationally efficient models for understanding the effects of different geometrical and topological properties on the fluid flow. Consequently, by patching Hele-Shaw cells, two new models were developed for simulating single-phase fluid through multiple smooth-walled fractures which provide solutions in less time comparing to NSE. The numerical models were solved within the COMSOL Multiphysics software package (COMSOL 2020). To investigate the accuracy of the proposed models, comparisons have been made with the NSE. The governing equations and the numerical method are summarized in Section 2, and the results of the investigations are discussed in Section 3.

## 2. Methodology
### 2.1. Governing Equations

The governing equations for incompressible Newtonian single-phase steady-state flow are the Navier Stokes Equations (NSE). Equations 1 and 2 represents a set of nonlinear partial differential equations for the velocity and pressure field (Zimmerman and Bodvarsson 1996). These equations can be expressed as momentum and mass conservation, respectively. Accordingly,

$$\frac{\partial \boldsymbol{u}}{\partial t} + (\boldsymbol{u}.\nabla)\boldsymbol{u} = -\frac{1}{\rho}\nabla P - \frac{\mu}{\rho}\nabla^2 \boldsymbol{u} \qquad \text{Eq. 1}$$

$$\nabla.\boldsymbol{u} = 0 \qquad \text{Eq. 2}$$

Equation 1 includes several terms: the time derivative of the fluid velocity, inertial forces, pressure forces and viscous forces respectively. In these equations, $\rho$ is the density $kg/m^3$, $\boldsymbol{u} = [u, v, w]$ is the velocity vector $m/s$, $P$ is the pressure $Pa$, and $\mu$ is the viscosity $Pa.s$ (Water in 20°C, $\rho = 1000\ kg/m^3$, $\mu = 0.001\ pa.s$).

As mentioned before, solving the NSE for the flow in a DFN is impractical; therefore, simplifications have to be applied. First of all, it is assumed that the flow is laminar and steady-state. Hence, the time derivative of velocity will be zero. Secondly, inertia forces are considered negligible in comparison with viscous and pressure forces. Lastly, flow is averaged across the fracture aperture and escribed by the velocity $\boldsymbol{u}_{[HS]} = [v, w]$. Therefore, the depth-averaging process transforms the computational domain for a three-dimensional domain with a 3-component velocity field into a two-dimensional computational domain, hence considerably reducing the computational requirement. Consequently, the governing equations for the flow in parallel plates separated by an aperture ($h$) can be simplified to the cubic-law, which is expressed by Equation 2. However, it fails to take into account the anisotropy of the fracture surface roughness. Improvement was made by applying the continuity equation into local velocities, which yields a Laplace equation (Equation 3).

$$\boldsymbol{u}_{[HS]} = -\frac{h^2}{12\mu}\nabla P \qquad \text{Eq. 2}$$

$$\nabla h \boldsymbol{u}_{[HS]} = 0 \qquad \text{Eq. 3}$$

The flow velocity of each fracture was approximated by $\underline{u}_{[HS]}$ in which $h$ is the fracture aperture. The Reynolds Number, $Re$, for flow in a fracture is defined by Equation 4.

$$Re = \frac{\rho U_{av} h}{\mu} \qquad \text{Eq. 4}$$

The average flow velocity along a fracture can be defined as:

$$U_{av} = \frac{Q}{A} \qquad \text{Eq. 5}$$

Where $A$ is the cross section area:

$$A = wh \qquad \text{Eq. 6}$$

By substituting Equation 5 into 4, we have Reynolds number for flow within fractures:

$$Re = \frac{\rho Q}{\mu w} \qquad \text{Eq. 7}$$

$w$ is the fracture depth perpendicular to the direction of the bulk flow. The hydraulic resistance $R$ ($Pa.s^{-1}.m^{-3}$) was calculated by Equation 8 (Oh et al. 2012).

$$R = \frac{\Delta P}{Q} \qquad \text{Eq. 8}$$

In this equation $\Delta P$ is the pressure difference between inlet and outlet, which is called global pressure difference($Pa$). The governing equation incorporates the local aperture profile into the solution, allowing for a more accurate description of the flow through a fracture where the aperture may be non-uniform. The HS equation is then solved numerically using the finite element method with several assumptions:

- In the Hele-Shaw model, fractures are represented as two-dimensional surfaces
- The fracture walls are smooth
- The flow is driven by pressure gradient
- The flow is Steady-state and Newtonian

## 2.2. Numerical methodology

The governing equations were solved using the COMSOL Multiphysics software package(COMSOL 2020). The single-phase laminar flow module and the coefficient from the boundary PDE module were used to solve the NSE (Equation 1) and the Hele-Shaw (HS) equation (Equation 3), respectively. For solving HS equation, the stationary steady-state solver and the MUMPS solver algorithm were employed. Moreover, the steady-state algebraic multigrid solver and the generalized minimal residual GMRES algorithm were utilized for solving the NSE. The mesh elements were discretized using second-order discretization for the velocity field and first-order discretization for pressure P2+P1 which was found to be accurate for the NSE models. Cubic elements with Lagrange shape functions were employed for discretization in the HS models. Fluid properties of water at 20°C were applied for both models.

## 2.3. Geometry of the computational model

To investigate the flow properties in a DFN, two representative models were developed and are presented in Figure 1. The first physical domain, labelled Model A, was defined as five fractures intersecting one another perpendicularly. The second model, called Model B, was defined as five fractures intersecting randomly. The geometrical properties are characterised by their aperture, width, length, intersection angle and intersection length ($h$, $w$, $L$, $\theta$ and $i$ respectively).

Several investigations were performed using these models. Firstly, the Hele-Shaw equation was tested by solving for the flow in Models A and B and comparing the results to the solution given by the Navier Stokes Equations (NSE). For this comparison, the apertures, intersection lengths, and intersection angles were held constant (0.02 $m$, 0.2 $m$, and 90° respectively). Next, the successfully validated Hele-Shaw model was used to investigate the influence of parameters like the intersection length and angle on the fluid flow within the fractures. This was achieved by varying the intersection length of the influent and effluent fractures (range of 0.2 $m$ – 0.45 $m$). Alternatively, the intersection angle was altered (range of 90° −65°). Another important factor in fracture network flow is the aperture of the fracture and its influence. Consequently, a last set of studies was performed by varying the aperture of Fracture 3 (range of 0.02 $m$ −0.09 $m$). When varying the intersection length, intersection angle, or the aperture, all other variables were kept constant. The dimension of all fractures in Model A was 1×1 $m$ .

The capability of the Hele-Shaw approximation to model randomly intersecting fractures was demonstrated by Model B. All studies conducted on Model B were done by assuming constant values for the geometrical properties. Model B constitutes of five fractures in series which was developed to demonstrate the ability of the method to simulate fluid flow in fractures with different apertures. In this model, the aperture of all fractures was set to 0.02 $m$, except for the second fracture with an aperture of 0.05 $m$. The width and length of the fractures were constant (1×1 $m$) as in Model A.

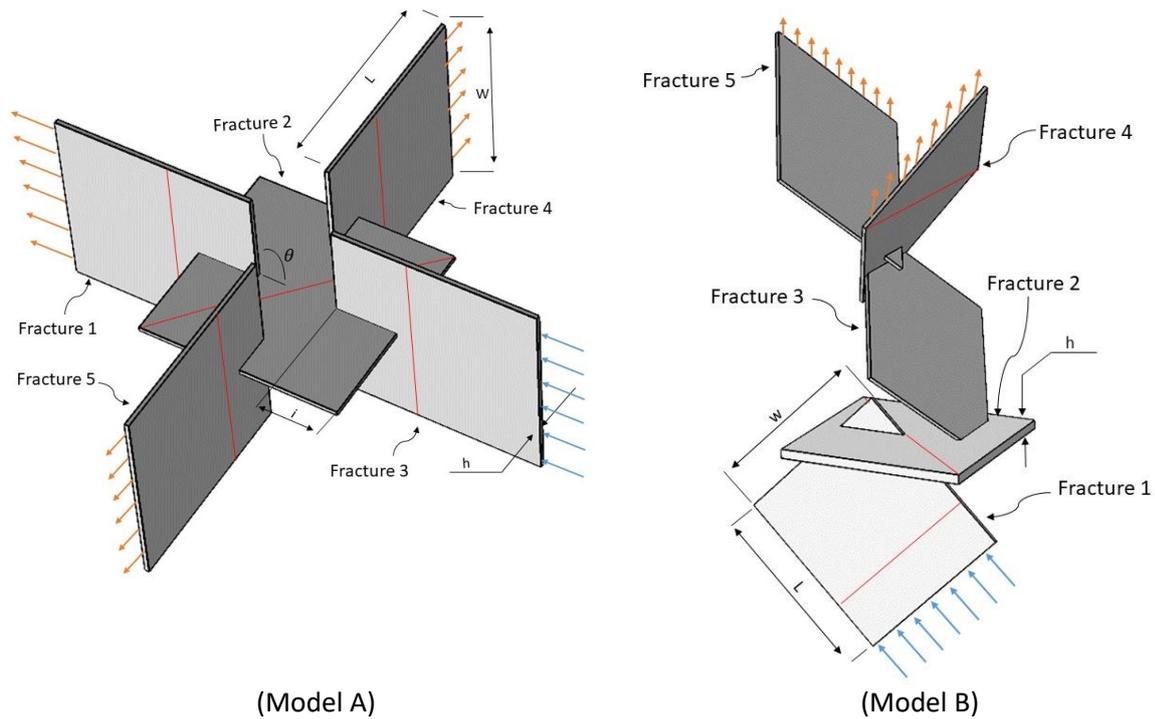

(Model A)                    (Model B)

Figure 1. Geometrical representation of Model A (perpendicularly intersecting) and B (randomly intersecting). The marked red lines indicate the location of specified cross sections for extracting and comparing data. The orange and blue arrows represent the pressure outlet and pressure inlets boundary conditions, respectively.

## 2.4. Boundary conditions

The boundary conditions were defined as follows. For the inlet and outlet boundaries, a constant pressure Dirichlet boundary conditions was applied over the edge of one of the fractures in the Hele-Shaw approximation, and over the surface area of a fracture in the NSE. The computational domain is depicted in Figure 1, with colour-coded boundary conditions. The blue colour represents the pressure inlet and the orange colour represents the pressure outlet. The pressure difference between the inlet and outlets is called the global pressure difference and varies depending on the study, but were always in the range of $0.005\ Pa - 0.0338\ Pa$ to keep the Reynolds number in the laminar regime.

A number of red lines are marked in Figure 1. These red lines are cross-sections from which representative data was extracted. The defined locations for the specified cross-sections in Model A are located at a distance of $0.7\ m$ from inlet and outlets. The main reason for choosing these locations was to study more precisely the effect of intersections on the fluid flow. In addition, a cross-section at mid-fracture in Model A was placed diagonally to investigate the fluid flow behaviour on both sides of the fracture. Similarly, the cross-sections in Model B are situated $0.2\ m$ from inlet in Fracture 1 and diagonally on fractures 1, 2 and 4.

## 2.5. Mesh

For numerical stability and mesh-independency, a mesh sensitivity study was conducted. For the sake of conciseness, the results of this study are not shown here. The optimal number of tetrahedral and prismatic elements to simulate the flow effectively by solving the Navier-Stokes Equations (NSE) was 288,379 elements. On the other hand, only 1632 triangular elements were needed for mesh-independent results using the Hele-Shaw approximation. It is important to note that the NSE are solved in a 3-dimensional domain, whereas the simplifications made in deriving the HS equation result in a 2-dimensional domain – the third dimension representing the direction of the aperture (the thickness of the fracture) is accounted for in the governing equation. For the HS simulations, the velocity and hydraulic had to be derived from the solved pressure gradient using Equation 2.

## 3. Results & Discussion

### 3.1. Validation of Hele-Shaw model

The ability and accuracy of the Hele-Shaw equations to model the flow in fracture networks was tested by comparing the results for the evaluated pressure and velocity fields – for Models A and B – with those given by solving full Navier-Stokes Equations (NSE) for the same network. These comparisons were carried out because the NSE is valid for subsurface fluid flow and is therefore the gold standard for validating simplified models (Wang et al. 2018; Zou et al. 2017a; Zou et al. 2017b). Because the Hele-Shaw model uses a 2D geometry and averages the flow across the width of the fracture in the governing equations, it is necessary to compare the results only in the centre plane of the NSE results. In other words, if we reflect the HS geometry on the NSE geometry, the 2D surfaces (fractures) of the HS must locate in the middle of the NSE's fractures. For instance, to compare obtained simulation results by solving the HS equation for flow through Fracture 2 of model A, the 2D surface must situate in the middle of NSE geometry perpendicular to the z-axis. The following results were conducted using a $0.01\ Pa$ pressure difference between the inlet and outlets. The evaluated Reynolds number for this pressure difference by NSE and HS are 2.91 and 2.95 respectively.

Comparisons of the velocity magnitude and pressure distribution were made by extracting data from specified cross-sections after solving the fluid flow using both the Hele-Shaw approximation and NSE. These results are presented in Figure 2. The original coordinate system (X, Y, Z) has the origin at the centre of the Model A. The X-axis is parallel to the flow of fractures 1, 2 and 3. Also, it represents the aperture of fractures 4 and 5. The Y-axis is in the flow direction of fractures 4 and 5. Similarly, it represents the aperture of fractures 1 and 2. The aperture of Fracture 2 is presented in Z-axis. Moreover, x is the coordinate of the fracture-local coordinate system. As can be seen in the figure, the fluid velocity is highest in the centre of the channel (x=0.5 $m$) than at the sides (x=0, x=1 $m$), which is the consequence of no-slip wall boundary conditions in NSE and zero flux boundary conditions in HS. The good agreement of the Hele-Shaw model with the NSE confirms the high accuracy of the Hele-Shaw approximation. Validation of Model B was conducted by the same approach, and the results are shown in Figure 3. Again, very good agreement is observed.

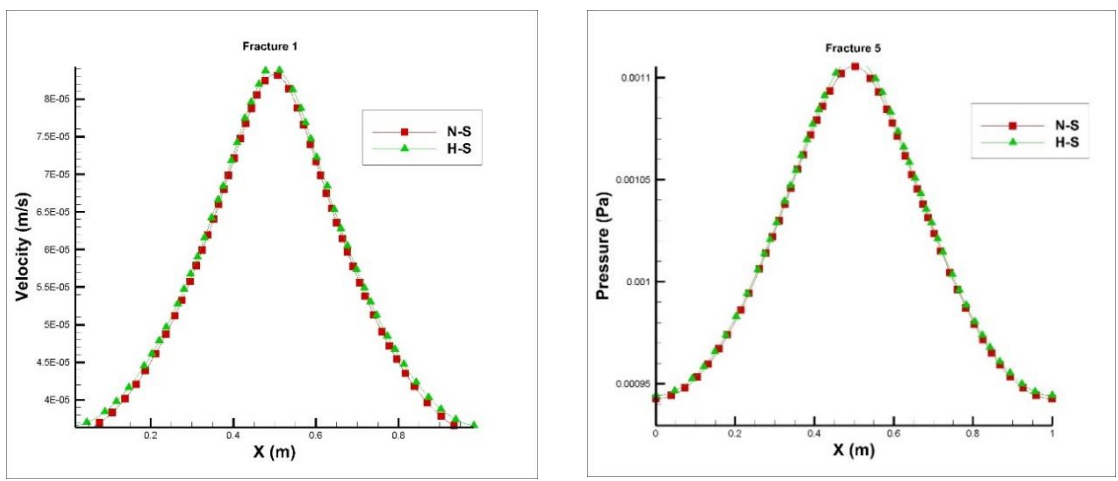

Velocity magnitude distribution on Fracture 1 cross section

Pressure distribution on fracture 5 cross section

(a)

(b)

Figure 2. Validation of Model A against Navier Stokes Equations. Pressure and velocity along the length (x) of Fracture 1 and Fracture 5 specified cross-sections

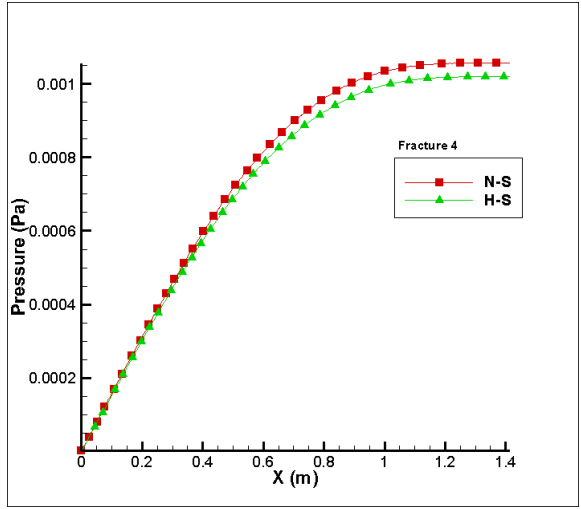
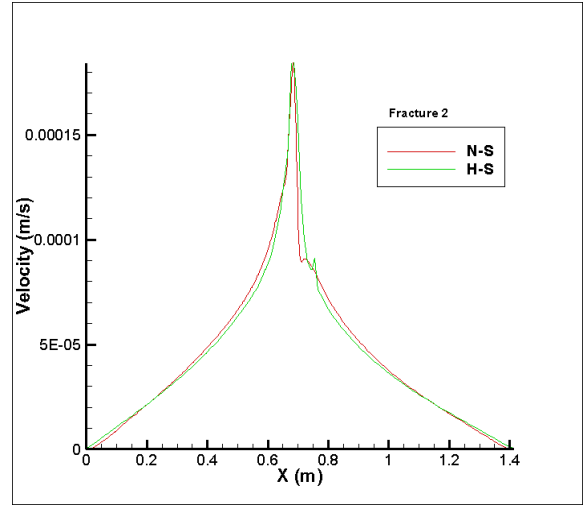

Pressure over Fracture 4 cross section

velocity magnitude over Fracture 2 cross section

(a)

(b)

Figure 3 – Validation of Model B against Navier Stokes Equations. Pressure and velocity along the length (x) of Fracture 4 and Fracture 2 specified cross-sections

In addition to the pressure and velocity fields, comparisons of the two methods were made by comparing the relationship between flow rate, hydraulic resistance and Reynolds number. The model was probed at Fracture 3 of Model A and Fracture 2 of Model B. The results are described in Tables 1 Table 2. These tables not only demonstrate the correlation between flow specifications, but they show the accuracy of the HS model. The relative error of the flow rate using the Hele-Shaw approximation is calculated as in Equation 9 below:

$$Error = \left[\frac{Q_{HS} - Q_{NS}}{Q_{HS}}\right] \times 100 \quad \text{Eq. 9}$$

As the global pressure difference increases, the hydraulic resistance has effectively remained constant. In addition, all the simulations performed found that the error from the Hele-Shaw model were less than 2%. The ability of the approach for simulating different arrangements of fractures was illustrated in Table 2 by connecting fractures in a random shape (Model B), which depicts the accuracy of the approach. All the presented parameters in Table 1 and 2 exhibit a similar trend.

Table 1. The results of the conducted Hele-Shaw approximation in comparison with Navier Stokes Equations (Model A)

| Global pressure difference ($Pa$) | Re | N-S flow rate ($m^3/s$) | H-S flow rate ($m^3/s$) | Error | Hydraulic resistance |
|---|---|---|---|---|---|
| **0.005** | 1.4 | 1.46E-06 | 1.48E-06 | 1% | 3.4334E+03 |
| **0.007** | 2.0 | 2.04E-06 | 2.07E-06 | 1% | 3.4342E+03 |
| **0.009** | 2.6 | 2.62E-06 | 2.66E-06 | 2% | 3.4356E+03 |
| **0.01** | 2.91 | 2.91E-06 | 2.96E-06 | 2% | 3.4364E+03 |
| **0.015** | 4.3 | 4.36E-06 | 4.43E-06 | 2% | 3.4408E+03 |
| **0.02** | 5.8 | 5.80E-06 | 5.91E-06 | 2% | 3.4461E+03 |
| **0.025** | 7.2 | 7.24E-06 | 7.39E-06 | 2% | 3.4524E+03 |
| **0.03** | 8.6 | 8.67E-06 | 8.87E-06 | 2% | 3.4595E+03 |
| **0.0338** | 9.7 | 9.75E-06 | 9.99E-06 | 2% | 3.4654E+03 |

Table 2. The results of the conducted Hele-Shaw approximation in comparison with Navier Stokes Equations (Model B)

| Global pressure difference ($Pa$) | Re | N-S flow rate $m^3/s$ | H-S flow rate $m^3/s$ | Error | Hydraulic resistance |
|---|---|---|---|---|---|
| **0.02** | 2.3 | 2.39E-06 | 2.33E-06 | 2% | 8.37E+03 |
| **0.03** | 3.5 | 3.57E-06 | 3.50E-06 | 2% | 8.41E+03 |
| **0.04** | 4.7 | 4.72E-06 | 4.67E-06 | 1% | 8.47E+03 |
| **0.05** | 5.8 | 5.87E-06 | 5.83E-06 | 1% | 8.52E+03 |
| **0.06** | 6.9 | 6.99E-06 | 6.998-06 | 1% | 8.58E+03 |
| **0.07** | 8.1 | 8.11E-06 | 8.16E-06 | 1% | 8.63E+03 |

Pressure distribution contours for Model B are shown in Figure 4. Contours A and B are shown for a global pressure differential of $0.01\ Pa$. These contours were obtained using the Hele-Shaw approximation and NSE, respectively. The pressure gradually decreases as the flow is forced toward the pressure outlet at the top. In this study, the longest time for solving the NSE with an ordinary desktop computer which was facilitated 32GB RAM and an i7-8700 CPU was 1 hour. In comparison, calculating the fluid flow using the HS approximation took less than 1 minute. This significant reduction in computational requirement is the key points of this approach which enable us to simulate the fluid flow in a large number of fractures time efficiently.

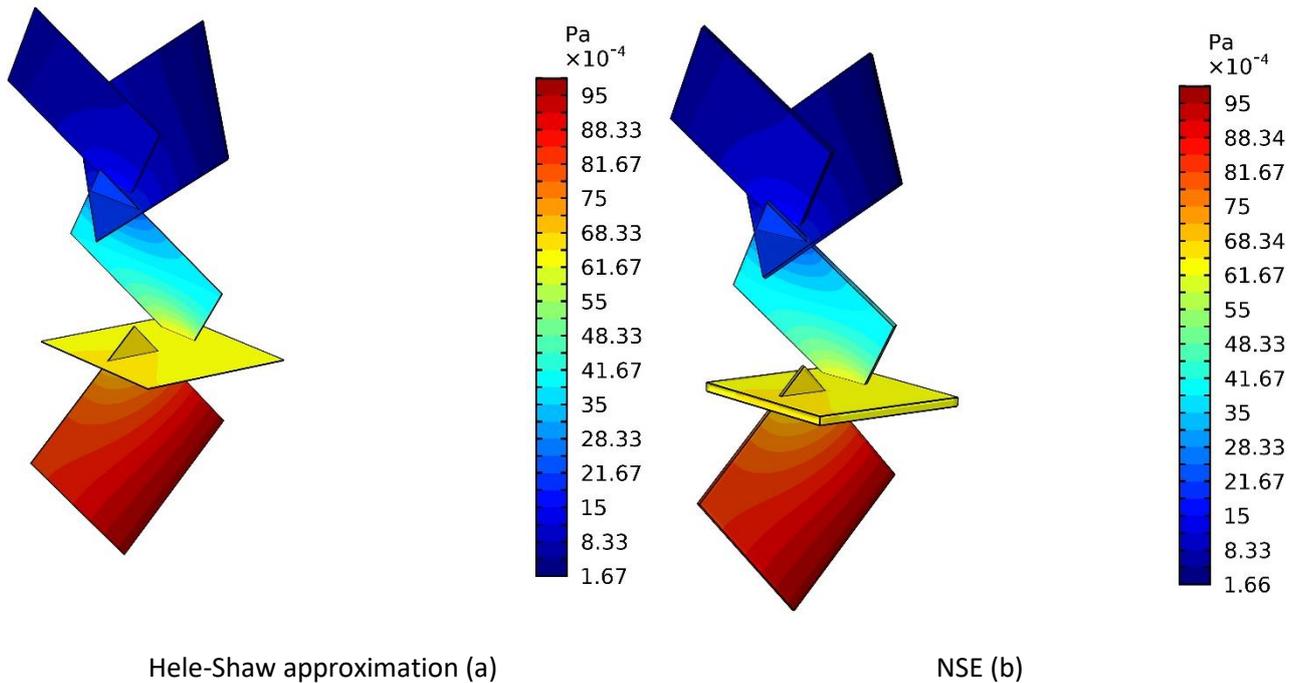

Hele-Shaw approximation (a)                  NSE (b)

Figure 4. Pressure distributions predicted by NSE and Hele-Shaw approximation for Model B

### 3.2. Influence of intersection length

Since the results of the Hele-Shaw model were found to be accurate, could the HS approximation can be used to investigate the influence of various geometric parameters on the flow within the fracture networks. In particular, the influence of intersection length, intersection angle and aperture size on fluid flow were studied using the HS approach. This was achieved by systematically adjusting the geometry of Model A. To investigate the influence of the intersection length, the positions of each fracture were modified so that every intersection length corresponded to the desired value.

Data was extracted from the cross-section of Fracture 1 in order to observe the impact of the intersection length. These results are shown in Figure 5. Increasing the intersection length increases the velocity and drops the pressure. With a larger conduit between fractures, there is less resistance to the flow resulting in higher velocities. Plotting the maximum velocity in the fracture against the intersection length, a quadratic relationship can be seen, as shown in Figure 6 b. However, the hydraulic resistance is negatively linearly correlated with the intersection length (see Figure 6 a).

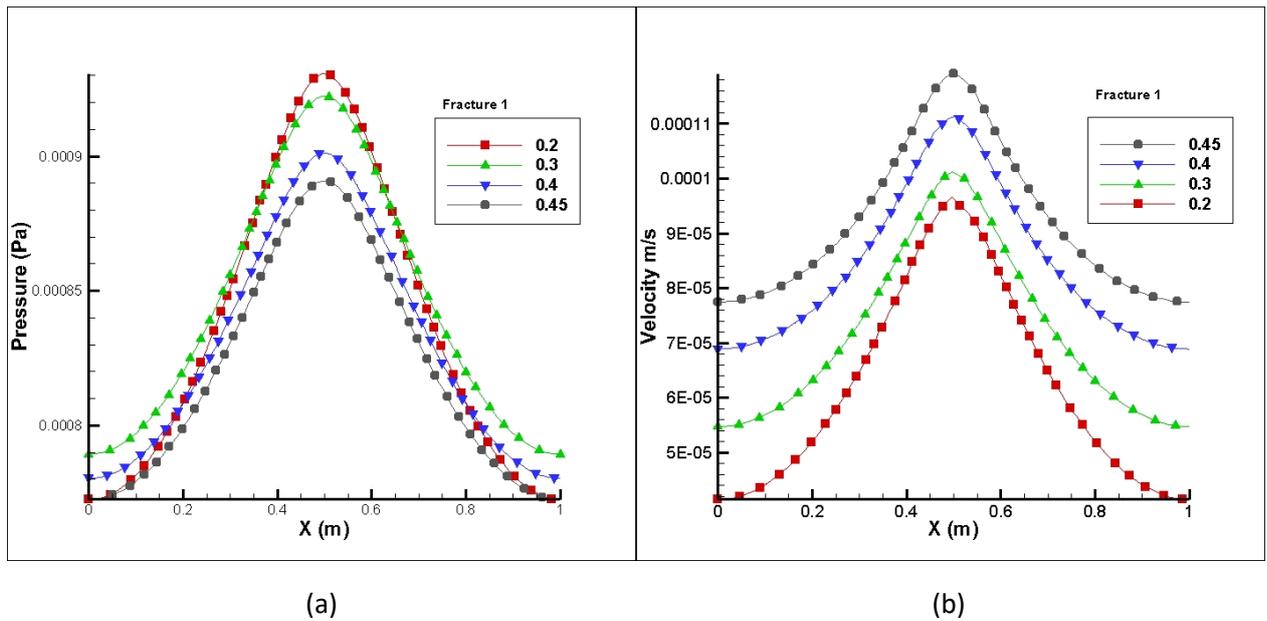

(a)                                       (b)

Figure 5 – The effect of different intersection length on velocity and pressure over the evaluated cross-section of Fracture 1 (Model A).

Hydraulic resistance, Reynolds Number, pressure difference and flow rate were evaluated for a range of $0.2\ m - 0.45\ m$ of intersection lengths and presented in Table 3. However, the flow rate – and consequently the Reynolds Number – increase as the intersection length increases. Since the intersection between fractures is the major chokepoint in the flow in fracture networks, it has a significant influence on the flow.

Table 3 – The effect of intersection length on flow rate of Fracture 3 (Model A)

| Intersection length ($m$) | Re | Flow rate $m^3/s$ | Global pressure difference ($Pa$) | Hydraulic resistance |
|---|---|---|---|---|
| **0.2** | 2.8 | 2.86E-06 | 0.01 | 3.50E+03 |
| **0.3** | 3.4 | 3.42E-06 | 0.01 | 2.92E+03 |
| **0.4** | 4.0 | 4.05E-06 | 0.01 | 2.47E+03 |
| **0.45** | 4.4 | 4.41E-06 | 0.01 | 2.27E+03 |

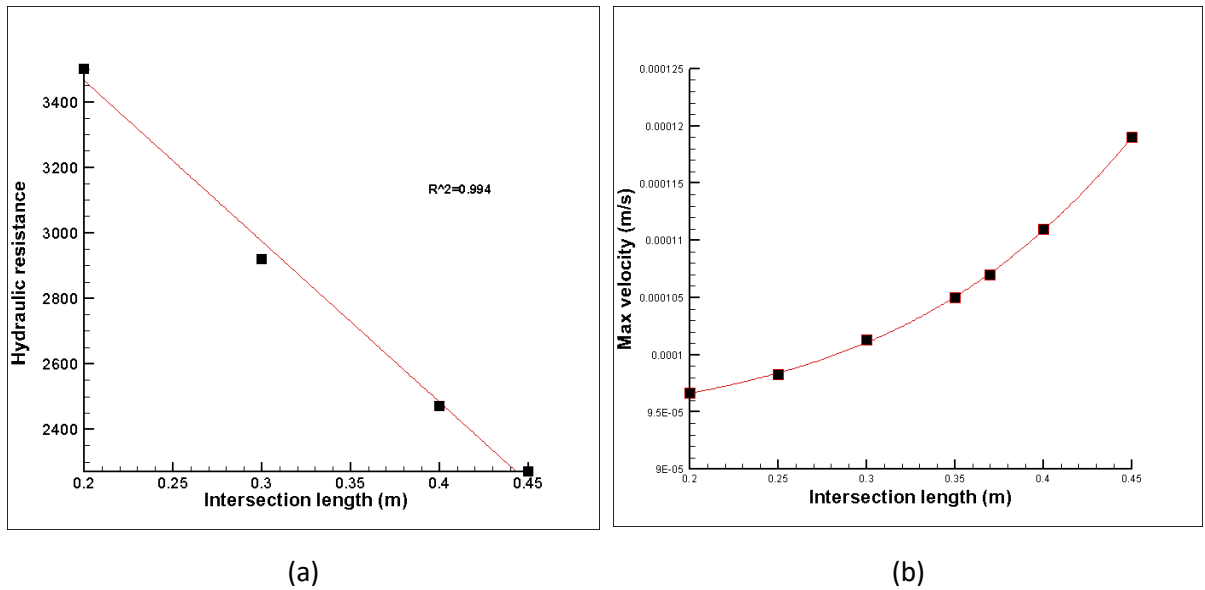

(a)  (b)

Figure 6 – The relationship between intersection length and hydraulic resistance of fracture 1 Model A (a). The correlation of intersection length and maximum velocity of Fracture 1 Model A (b)

### 3.3. The effect of different intersection angles

As shown in Figure 1, the intersection angle, $\theta$, was varied from 90° to 65°. For more clarification physical geometry of Model A with $\theta = 65°$ is presented in Figure 7. Simulations for the cases of $\theta = 90°, 80°, 75°, 70°$ and $65°$ were performed, and their results summarized in Table 4, which are in accordance with results of a 2019 paper by Bo li et al. (Li et al. 2019). Reducing the fracture angle to less than 65° was not possible because perpendicular fractures to midplane can become connected to each other.

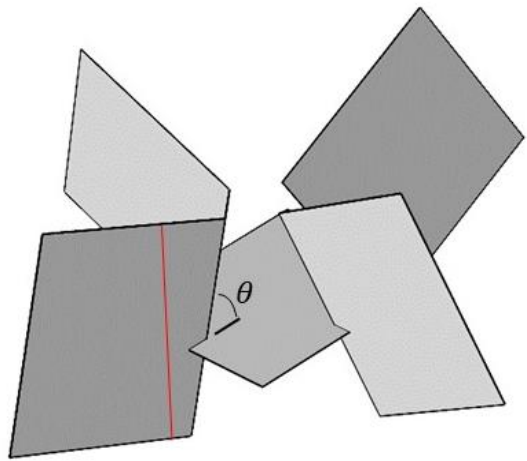

Figure 7 – *Geometrical representation of Model A – $\theta = 65°$*

When the flow has sufficiently large velocity, the flow rate can be affected intensely by intersection angle and fracture shape (Liu et al. 2020), but the present results show that increasing the angle of the fractures in the linear regime of fluid flow results in negligible changes. The results of different intersection angles at the specified cross sections are presented in Figure 8 and Table 4. The pressure distribution and velocity magnitude along the probed cross-section of Fracture 1 experienced only small changes by applying different intersection angles.

Table 4 – Effect of different intersection angles on Model A

| Intersection angle | Intersection length $(m)$ | Flow rate $m^3/s$ | Re | Global pressure difference $(Pa)$ | Hydraulic resistance |
|---|---|---|---|---|---|
| **90°** | 0.1 | 2.27E-06 | 2.2 | 0.01 | 4.40E+03 |
| **80°** | 0.1 | 2.29E-06 | 2.2 | 0.01 | 4.37E+03 |
| **75°** | 0.1 | 2.28E-06 | 2.2 | 0.01 | 4.38E+03 |
| **70°** | 0.1 | 2.28E-06 | 2.2 | 0.01 | 4.38E+03 |
| **65°** | 0.1 | 2.28E-06 | 2.2 | 0.01 | 4.39E+03 |

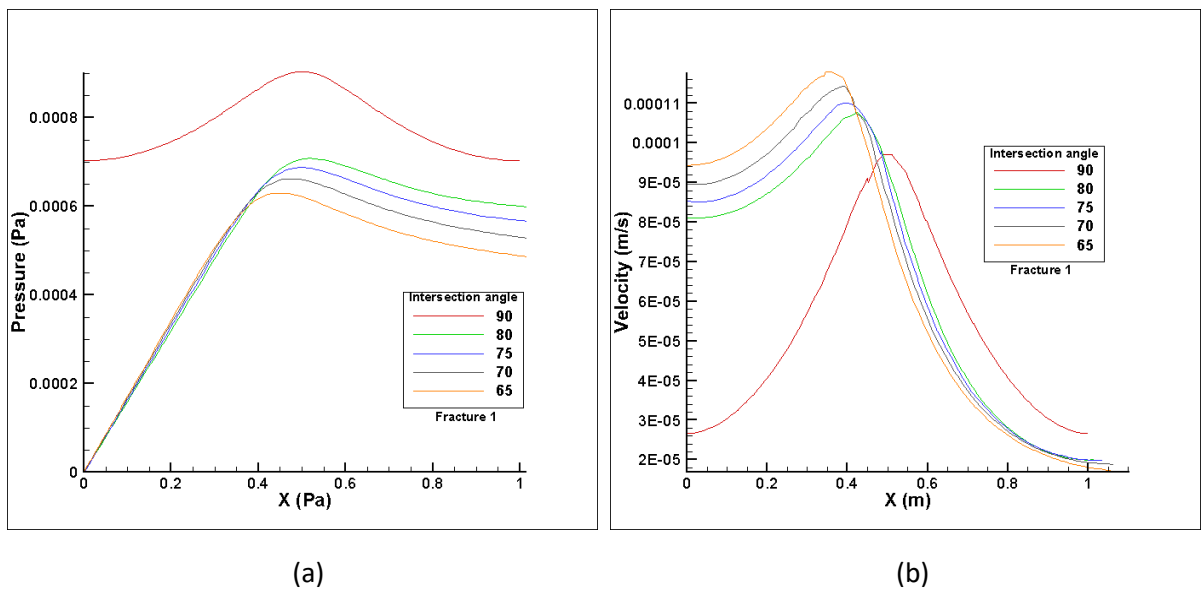

(a)            (b)

Figure 8 –The effect of different intersection angles on pressure and velocity over the length of Fracture 1 (Model A).

### 3.4. The effect of different fracture apertures

The fracture aperture was changed to study the effect on fluid flow. The minimum and maximum chosen fracture sizes for this simulation were 0.02 $m$ and 0.09 $m$. Figure 9 plots the pressure and velocity on the defined cross sections for different aperture sizes. As expected, increasing the fracture aperture caused the velocity and local pressure to drop because the global pressure gradient is now acting over a larger cross-sectional area. When increasing the aperture from 0.02 $m$ to 0.05$m$, the local pressure at a characteristic cross-section is reduced significantly (see Figure 9 a). However, when the aperture is increased further the change is much smaller. This indicates that, the pressure is more sensitive to fractures with smaller apertures. As presented in Table 5 with a larger fracture aperture, there is less resistance to the flow resulting in higher velocities.

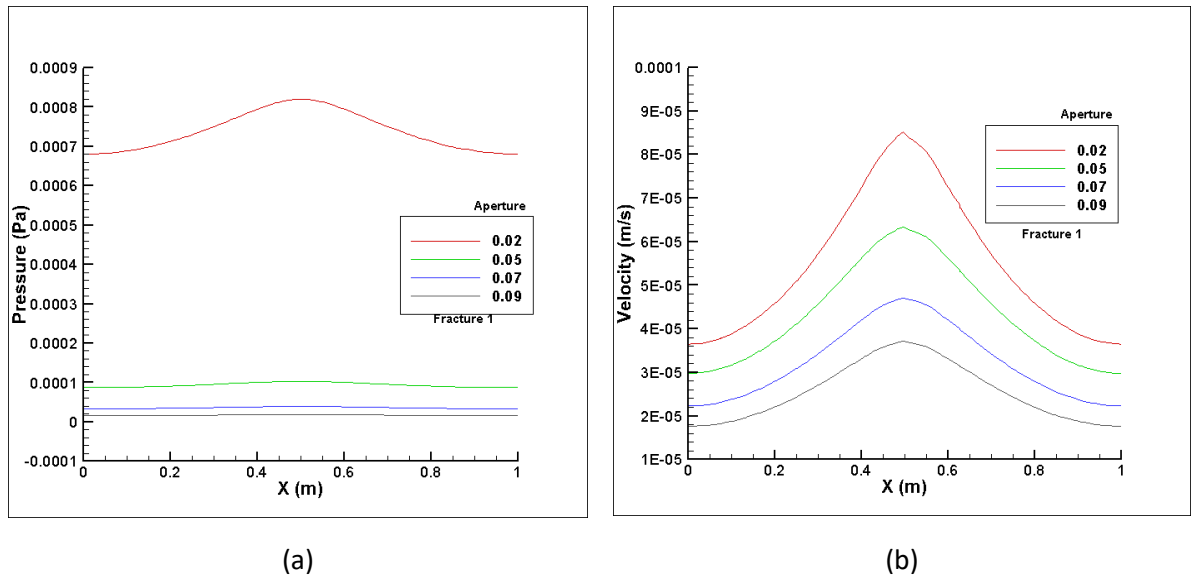

(a)                  (b)

Figure 9 –The effect of different apertures on pressure and velocity over the length of Fracture 1 (Model A).

Table 5– Effect of different apertures on Model A

| Aperture ($m$) | Re | Flow rate $m^3/s$ | Global pressure difference $Pa$ | Hydraulic resistance |
|---|---|---|---|---|
| **0.02** | 3.0 | 2.96E-06 | 0.01 | 3.38E+03 |
| **0.05** | 5.9 | 5.87E-06 | 0.01 | 1.70E+03 |
| **0.07** | 6.2 | 6.15E-06 | 0.01 | 1.63E+03 |
| **0.09** | 6.2 | 6.24E-06 | 0.01 | 1.60E+03 |

## 4. Conclusion

A computationally efficient numerical model of fracture networks by connecting Hele-Shaw cells was conducted to explore the effects of various intersection parameters on the hydraulic resistance of the fluid flow. The results obtained using Hele-Shaw patching method and Navier Stokes equations were compared to illustrate the validity of the Hele-Shaw approximation for geothermal applications which is significantly computational efficient. The Hele-Shaw model achieved accurate results at low Reynolds numbers with less than 2% Error in 1 minute, while solving Navier Stokes equations took more than an hour for same physical geometries. The key point of this research is this significant reduction in computational requirements, which can provide accurate solutions for large scale problems. A structural model with five fractures was used as the baseline model for accuracy evaluation of the Hele-Shaw patching method (Model A). Since the results of the Hele-Shaw model were found to be accurate, it was used to investigate the effects of various geometrical and topological parameters on the hydraulic resistance. Specifically, the Model B – which employs random intersections of the fractures – was developed to demonstrate the capability of this approach for simulating random geometries as potentially seen in nature. Results show an approximately linear relationship between the hydraulic resistance and the intersection length changes. On one hand, increasing the intersection length causes slight growth in velocity and flow rate. On the other, it causes a moderate drop in hydraulic resistance. In addition, increasing the intersection angle causes negligible changes in flow rate and hydraulic resistance. The study also investigates the effect of aperture on the flow in a fracture network with five fractures. The result shows an inverse relationship between the fracture aperture and the hydraulic resistance. Increase in fracture aperture leads to a decrease in hydraulic resistance. This is intuitively correct as larger apertures provides a bigger intersection area for the fluid to flow. For a given pressure drop, this results in a reduction of velocity and pressure over the defined cross-sections. Further investigation on five fractures in series re-emphasize the linearity in the effect of intersection length to the hydraulic resistance.